\documentclass[11pt]{article}\textwidth 6.5in\textheight 9in
\usepackage{amssymb}\usepackage[colorlinks]{hyperref}\usepackage{color}
	\usepackage{stmaryrd}\usepackage{mathrsfs}
	\usepackage{graphicx}
	\usepackage{amsmath}
\usepackage{caption}

\begin{document}
\setlength{\captionmargin}{27pt}
\newcommand\hreff[1]{\href {http://#1} {\small http://#1}}
\newcommand\trm[1]{{\bf\em #1}} \newcommand\emm[1]{{\ensuremath{#1}}}
\newcommand\prf{\paragraph{Proof.}}\newcommand\qed{\hfill\emm\blacksquare}

\setcounter{tocdepth}{3} 

\newtheorem{thr}{Theorem} 
\newtheorem{lmm}{Lemma}
\newtheorem{cor}{Corollary}
\newtheorem{con}{Conjecture} 
\newtheorem{prp}{Proposition}

\newtheorem{blk}{Block}
\newtheorem{dff}{Definition}
\newtheorem{asm}{Assumption}
\newtheorem{rmk}{Remark}
\newtheorem{clm}{Claim}
\newtheorem{example}{Example}

\newcommand{\ab}{a\!b}
\newcommand{\yx}{y\!x}
\newcommand{\yux}{y\!\underline{x}}

\newcommand\floor[1]{{\lfloor#1\rfloor}}\newcommand\ceil[1]{{\lceil#1\rceil}}

\newcommand{\lea}{<^+}
\newcommand{\gea}{>^+}
\newcommand{\eqa}{=^+}

\newcommand\om{\overline{\mu}}
\newcommand\on{\overline{\nu}}

\newcommand{\lel}{<^{\log}}
\newcommand{\gel}{>^{\log}}
\newcommand{\eql}{=^{\log}}

\newcommand{\lem}{\stackrel{\ast}{<}}
\newcommand{\gem}{\stackrel{\ast}{>}}
\newcommand{\eqm}{\stackrel{\ast}{=}}

\newcommand\edf{{\,\stackrel{\mbox{\tiny def}}=\,}}
\newcommand\edl{{\,\stackrel{\mbox{\tiny def}}\leq\,}}
\newcommand\then{\Rightarrow}

\newcommand\km{{\mathbf {km}}}\renewcommand\t{{\mathbf {t}}}
\newcommand\KM{{\mathbf {KM}}}\newcommand\m{{\mathbf {m}}}
\newcommand\md{{\mathbf {m}_{\mathbf{d}}}}\newcommand\mT{{\mathbf {m}_{\mathbf{T}}}}
\newcommand\K{{\mathbf K}} \newcommand\I{{\mathbf I}}

\newcommand\II{\hat{\mathbf I}}
\newcommand\Kd{{\mathbf{Kd}}} \newcommand\KT{{\mathbf{KT}}} 
\renewcommand\d{{\mathbf d}} 
\newcommand\D{{\mathbf D}}
\renewcommand\H{\mathbf{H}}

\newcommand\w{{\mathbf w}}
\newcommand\Ks{\mathbf{Ks}} \newcommand\q{{\mathbf q}}
\newcommand\E{{\mathbf E}} \newcommand\St{{\mathbf S}}
\newcommand\M{{\mathbf M}}\newcommand\Q{{\mathbf Q}}
\newcommand\ch{{\mathcal H}} \renewcommand\l{\tau}
\newcommand\tb{{\mathbf t}} \renewcommand\L{{\mathbf L}}
\newcommand\bb{{\mathbf {bb}}}\newcommand\Km{{\mathbf {Km}}}
\renewcommand\q{{\mathbf q}}\newcommand\J{{\mathbf J}}
\newcommand\z{\mathbf{z}}

\newcommand\B{\mathbf{bb}}\newcommand\f{\mathbf{f}}
\newcommand\hd{\mathbf{0'}} \newcommand\T{{\mathbf T}}
\newcommand\R{\mathbb{R}}\renewcommand\Q{\mathbb{Q}}
\newcommand\N{\mathbb{N}}\newcommand\BT{\{0,1\}}
\newcommand\FS{\BT^*}\newcommand\IS{\BT^\infty}
\newcommand\FIS{\BT^{*\infty}}\newcommand\C{\mathcal{L}}
\renewcommand\S{\mathcal{C}}\newcommand\ST{\mathcal{S}}
\newcommand\UM{\nu_0}\newcommand\EN{\mathcal{W}}

\newcommand{\supp}{\mathrm{Supp}}

\newcommand\lenum{\lbrack\!\lbrack}
\newcommand\renum{\rbrack\!\rbrack}
\renewcommand\T{\mathbf{T}}
\renewcommand\qed{\hfill\emm\square}

\title{\vspace*{-3pc} The Randomness Deficiency Function and the Shift Operator}

\author {Samuel Epstein\footnote{JP Theory Group. samepst@jptheorygroup.org}}

\maketitle
\begin{abstract}
	Almost surely, the difference between the randomness deficiencies of two infinite sequences will be unbounded with respect to repeated iterations of the shift operator. 
\end{abstract}
\section{Introduction}
In \cite{EpsteinAlgPhysics23}, a result was proven about thermodynamics and product spaces. It was shown that all typical states of product spaces cannot have their marginal algorithmic thermodynamic entropies in synch during the course computable ergodic dynamics. This result was over all computable metric spaces, using the foundation of \cite{HoyrupRo09}. This paper shows the special case of the Cantor space and the shift operator, which could be of independent interest from algorithmic physics. It is proved using the uniform measure, but with a little bit of work, it can be generalized to two different computable probability measures.

The result is as follows. Let $\K$ be the prefix free Kolmogorov complexity. Let $\m$ be the algorithmic probability and $\I(x:y)=\K(x)+\K(y)-\K(x,y)$ be the mutual information term. The mutual information between two infinite sequences \cite{Levin74} is $\I(\alpha:\beta)=\log\sum_{x,y\in\FS}\m(x|\alpha)\m(y|\beta)2^{\I(x:y)}$. The halting sequence is $\ch\in\IS$. The shift operator is $\sigma$, where $\sigma(\alpha_1\alpha_2\alpha_3\dots)=\alpha_2\alpha_3\dots$. The uniform measure over $\IS$ is $\lambda$. The randomness deficiency of $\alpha\in\IS$ is $\D(\alpha)=\sup_n\left(n-\K(\alpha[0..n])\right)$.  For infinite sequences $\alpha,\beta\in\IS$, $(\alpha,\beta)$ encodes them with alternating bits.\\

\noindent\textbf{Theorem.}\textit{
	\begin{enumerate}
		\item[(a)] If $(\alpha,\beta)$ is ML Random and $\I((\alpha,\beta):\ch)<\infty$ then $\sup_n |\D\left(\sigma^{(n)}\alpha\right)-\D\left(\sigma^{(n)}\beta\right)|=\infty$.
		\item[(b)] For $\lambda\times \lambda$ almost surely, $\sup_n |\D\left(\sigma^{(n)}\alpha\right)-\D\left(\sigma^{(n)}\beta\right)|=\infty$.
	\end{enumerate}}
\section{Conventions}
For positive real functions $f$, by ${\lea}f$, ${\gea}f$, ${\eqa}f$, and ${\lel} f$, ${\gel} f$, ${\sim} f$ we denote ${\leq}\,f{+}O(1)$, ${\geq}\,f{-}O(1)$, ${=}\,f{\pm}O(1)$ and ${\leq}\,f{+}O(\log(f{+}1))$, ${\geq}f\,{-}O(\log(f{+}1))$, ${=}\,f{\pm}O(\log(f{+}1))$. Furthermore, ${\lem}f$, ${\gem}f$ denotes $< O(1)f$ and $>f/O(1)$. The term and ${\eqm}f$ is used to denote ${\gem}f$ and ${\lem}f$.


For a universal lower computable $P$ test $\t$ over $\IS$, $\log \t(\alpha) \eqa \D(\alpha|P)$. The function $\T$ is a universal lower computable $\lambda\times\lambda$ test, where if $T$ is a lower computable $\lambda\times\lambda$ test,  with $\int Td\lambda\times d\lambda<1$, then $T\lem \T$. If $\T(\alpha,\beta)=\infty$, then $(\alpha,\beta)$ is not ML random.

\begin{prp}
\label{prp:cons}
For partial computable $f:\IS\rightarrow\IS$,
    $\I(f(\alpha):\ch)\lea \I(\alpha:\ch)+\K(f)$.
\end{prp}
\begin{thr}[\cite{EpsteinAnomalies24}]
\label{thr:set}
    For computable probability $P$ over $\IS$, $Z\subset\IS$, $|S|=2^s$, $s\lel \max_{\alpha\in Z}\D(\alpha|P)+\I(\langle Z\rangle:\ch)+O(\log \K(P))$.
\end{thr}

\begin{thr}[\cite{Vereshchagin21,Levin74,Geiger12}]
	\label{thr:coninfo}
	$\Pr_{\mu}(\I(\alpha:\ch)>n)\lem 2^{-n+\K(\mu)}$.
\end{thr}

\section{Results}
\begin{thr} $ $\\
	\vspace*{-0.5cm}
		\begin{enumerate}
		\item[(a)] If $(\alpha,\beta)$ is ML Random and $\I((\alpha,\beta):\ch)<\infty$ then $\sup_n |\D(\sigma^{(n)}\alpha)-\D(\sigma^{(n)}\beta)|=\infty$.
		\item[(b)] For $\lambda\times \lambda$ a. s.,  $\sup_n |\D\left(\sigma^{(n)}\alpha\right)-\D\left(\sigma^{(n)}\beta\right)|=\infty$.
	\end{enumerate}
\end{thr}
	\begin{prf}
	(a)   Assume $\I((\alpha,\beta):\ch)<\infty$ and $c=\sup_{t\in\N}|\D(\sigma^{(t)}\alpha|P)-\D(\sigma^{(t)}\beta|P)|$. Let $Z_n=\{ \sigma^{(t)}\alpha : t \in \{1,\dots, 2^n\} \}$, where $\K(Z_n|(\alpha,\beta))\lea \K(n)$. Due to Theorem \ref{thr:set} and Proposition \ref{prp:cons},
     \begin{align*}
     n &< \max_{\gamma \in Z_n} \log \t(\gamma) + \I(Z_n:\ch)\\
     &< \max_{\gamma \in Z_n} \log \t(\gamma) + \I((\alpha,\beta):\ch)+O(\K(n))\\
     &< \max_{\gamma \in Z_n} \log \t_\mu(\gamma)+O(\K(n)).
\end{align*}
Let $t_n = \arg\max_{t\in\{1,\dots,2^n\}}\t(\sigma^{(t)}\alpha)$. So $\K(t_n)\lea n+\K(n)$.
Let $V_n = \{\gamma : \t(\gamma)>2^{n-O(\log n)}\}$ and $W_n =\{\gamma : \t(\gamma)>2^{n-O(\log n)-c}\}$. One can define the $\lambda\times \lambda$ test $T_n(\gamma,\xi) = [\gamma \in \sigma^{-(t_n)}(W_n)\textrm{ and } \xi \in \sigma^{(-t_n)}(V_n)]2^{2n-O(\log n)-c}$. Thus 
$$\T(\alpha,\beta)\gem \sum_n\m(T_n)T_n(\alpha,\beta) \gem \sum_n 2^{n-O(\log n) -c}>\infty.$$ 
So $\D(\alpha,\beta)=\infty$ and $(\alpha,\beta)$ is not ML Random.	\\	
	
	(b) By the results of (a), if $\sup_n |\D\left(\sigma^{(n)}\alpha\right)-\D\left(\sigma^n(\beta)\right)|<\infty$, then $(\alpha,\beta)$ is not ML random or $\I((\alpha,\beta):\ch)=\infty$. By Theorem \ref{thr:coninfo}, $\lambda\{\gamma:\I(\gamma:\ch)=\infty\}=0$. Thus $(\alpha,\beta)$ is in a $\lambda$ null set, and thus also a $\lambda\times\lambda$ null set.	\qed
\end{prf}

\end{document}